\documentclass[fleqn,usenatbib]{mnras}

\usepackage{newtxtext,newtxmath}

\usepackage[T1]{fontenc}
\usepackage{ae,aecompl}

\usepackage{graphicx}	
\usepackage{amsmath}	
\usepackage{amssymb}	
\usepackage{natbib}
\usepackage{mathrsfs}

\title[Phased array feed survey]{A 21~cm pilot survey for pulsars and transients using the Focal L-Band Array for the Green Bank Telescope}

\author[K. M. Rajwade et al.]{\noindent 
K. M. Rajwade,$^{1}$\thanks{E-mail: kaustubh.rajwade@manchester.ac.uk}
D. Agarwal,$^{2,3}$
D. R. Lorimer,$^{2,3,4}$
N. M. Pingel,$^{2,3,7}$
D. J. Pisano,$^{2,3,4}$ 
\newauthor M. Ruzindana,$^{5}$
B. Jeffs,$^{5}$
K. F. Warnick,$^{5}$
D. A. Roshi$^{6}$
and M. A. McLaughlin$^{2,3,4}$
\\
$^{1}$Jodrell Bank Centre for Astrophysics, University of Manchester, Oxford Road, Manchester M13 9PL, UK\\
$^{2}$Department of Physics and Astronomy, West Virginia University, Morgantown, WV 26506,USA\\
$^{3}$Center for Gravitational Waves and Cosmology, Chestnut Ridge Research Building, West Virginia University, Morgantown, WV 26506, USA \\
$^{4}$ Adjunct Astronomer, Green Bank Observatory, Green Bank, WV 24944, USA \\
$^{5}$ Department of Electrical and Computer Engineering, Brigham Young University, Provo, UT, USA \\
$^{6}$ Arecibo Observatory, HC-3, Box 53995, Arecibo, PR 00612, USA \\
$^{7}$ Research School of Astronomy and Astrophysics, Australian National University, Canberra, ACT 2611, Australia
}

\date{Accepted XXX. Received YYY; in original form ZZZ}

\pubyear{2015}

\begin{document}
\label{firstpage}
\pagerange{\pageref{firstpage}--\pageref{lastpage}}
\maketitle

\begin{abstract}
Phased Array Feed (PAF) receivers are at the forefront of modern day radio astronomy. PAFs are currently being developed for spectral line and radio continuum surveys and to search for  pulsars and fast radio bursts. Here, we present results of the pilot survey for pulsars and fast radio bursts using the Focal plane L-band Array for the Green Bank Telescope (FLAG) receiver operating in the frequency range of 1.3--1.5~GHz. With a system temperature of $\sim$18~K, the receiver provided unprecedented sensitivity to the survey over an instantaneous field of view (FoV) of 0.1 deg$^{2}$. For the survey, we implemented both time and frequency domain search pipelines designed to find pulsars and fast radio bursts that were validated by test pulsar observations. Although no new sources were found, we were able to demonstrate the capability of this instrument from observations of known pulsars. We report an upper limit on the rate of fast radio bursts above a fluence of 0.36~Jy ms to be 1.3 $\times$ 10$^6$ events per day per sky. Using population simulations, we show that the FLAG will find a factor of 2--3 more pulsars in same survey duration compared to its single pixel counterpart at the Green Bank Telescope. We also demonstrate that the new phased array receiver, ALPACA for the Arecibo telescope, will be a superior survey instrument and will find pulsars at a higher rate than most contemporary receivers by a factor of 2--10.
\end{abstract}




\section{Introduction}

Since the discovery of pulsars over 50 years ago \citep{hpb+68}, extensive radio surveys have been performed to search for pulsating neutron stars. Currently, almost 3000 radio pulsars have been found by radio surveys of the Galaxy~\citep{ke16}, its globular cluster systems \citep{ran2008} and the Magellanic Clouds~\citep{cr2001}. Population studies have shown the total number of active radio pulsars to be of order $10^5$~\citep[see, e.g.,][]{fk06,kar17}. As a result, we have only sampled a very small fraction of the entire population of pulsars. Moreover, the discovery of Fast Radio Bursts (FRBs) and the rapid growth of this field in the last decade~\citep{lo07,th13,pe15,ca17,cha17} has given even further impetus to surveys for these and other radio transients. 

Among the pulsar population are the millisecond pulsars (MSPs) that are characterised by their millisecond rotation periods, smaller magnetic fields and smaller light cylinders compared to their long-period (normal pulsar) counterparts~\citep{ku1994}. Currently, MSPs are being extensively used in Pulsar Timing Array (PTA) projects as they are extremely stable timers~\citep{mcl13}. More MSP discoveries will improve angular sampling of sources in the sky thus, decreasing the time of detecting gravitational waves from PTAs~\citep{kr2016}. This has exacerbated the need for discovering more MSPs in current, and future pulsar surveys since only a few hundred have been discovered so far from approximately 3$\times$10$^{4}$ expected MSPs in the Galaxy~\citep{le13}.

FRBs are millisecond duration, bright radio flashes that occur over the entire sky~\citep{lo07}. Their high dispersion measures compared to the ones measured for Galactic sources confirm their extra-galactic origin. Currently about 64 FRBs have been published ~\citep{lo07,th13,ch16,pe15b,ca17,sh2018,chime2019a} and an up-to-date catalogue can be found online~\citep{pet16}\footnote{http://www.frbcat.org}. Though a number of these sources have been discovered, their origin is still a matter of debate. The discovery of only two repeating FRBs along with the localization and identification of a host galaxy for one of them has underscored the importance in finding these objects~\citep{cha17,te17,chime2019b}. Thus, the focus has shifted to fast, large field of view radio surveys with real-time searches for FRBs and other fast transients. Among the many surveys being carried out, many of the major radio facilities are carrying out experiments of this nature: the Canadian HI Intensity Mapping Experiment~\citep[CHIME;][]{CHIME2018}; the Green Bank Telescope~\citep[GREENBURST;][]{greenburst};
the Upgraded Molonglo Synthesis Radio Telescope~\citep[UTMOST;][]{ca16};
and the Australian Square Kilometre Array Pathfinder~\citep[ASKAP;][]{jam18}. 
\newline
Alternative methods to achieve large FoVs with greater sensitivities is to employ PAFs for surveys with large single dish radio telescopes. Recent developments in antenna feeds, and instrumentation has led to the creation of PAFs on all the major radio telescopes around the world.~\cite{deng2017} and~\cite{mal2018} have presented results from PAFs employed on the Parkes 64-m dish and the 100-m Effelsberg radio telescope respectively that show PAFs to be promising survey instruments. An on-going project to build a 40-beam PAF for the Arecibo telescope called the Advanced L-band cryogenic Phased Array Camera for Arecibo(ALPACA) is expected to improve the survey capabilities of the telescope~\footnote{\url{https://nsf.gov/awardsearch/showAward?AWD_ID=1636645&HistoricalAwards=false}}~\citep{cortes2016}. The Focal L-Band Array for the Green Bank Telescope (FLAG) is one such PAF that has been built for the Green Bank Telescope (GBT). It is one of the most sensitive PAFs to date and will result in a three to five-fold increase in survey speeds for HI mapping and transients, including arc-minute localisation of fast radio transients and robust determination of fluence for detections that are made with multiple beams. In this paper, we present the results of the first pilot survey for pulsars and FRBs using the FLAG. The survey description is presented in Section 2. The search techniques are presented in Section 3. In Section 4, we compute the limits on FRB rates. We discuss the implications from the present and future surveys with PAFs at different telescopes in Section 5. Our conclusions are presented in Section 6.

\section{Survey Description}
\subsection{FLAG receiver}

\begin{figure*}
\centering
\includegraphics[scale = 0.4]{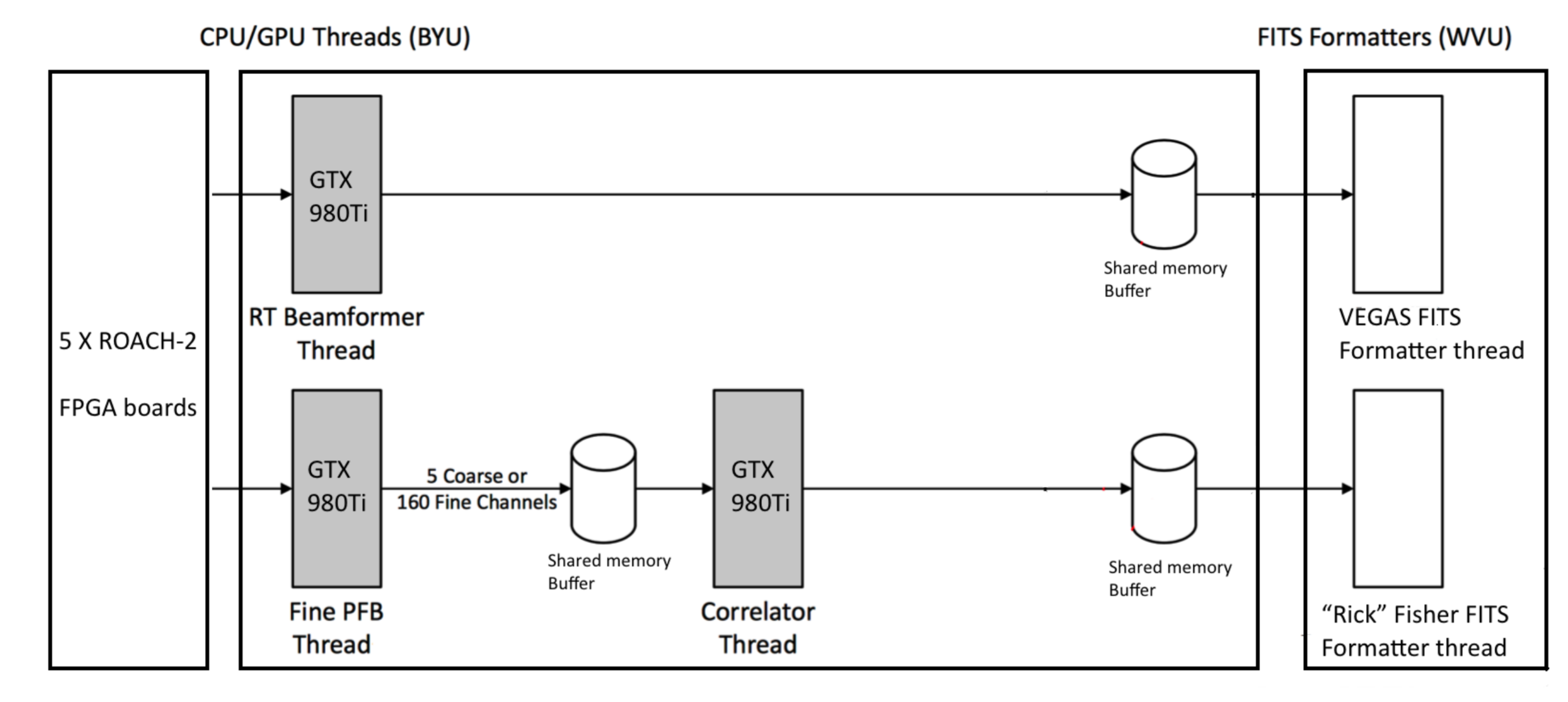}
\caption{Flow diagram representing the basic signal path of the FLAG processing pipeline. Some aspects of the figure have been adopted from Fig.~1 in~\protect\cite{raj2018} }
\label{fig:pipeline}
\end{figure*}

The FLAG project is a joint collaboration between Green Bank Observatory (GBO), West Virginia University (WVU), Brigham Young University (BYU) and the National Radio Astronomy Observatory (NRAO)~\citep[See][for more details]{roshi2019}. The first commissioning of the PAF was done in the summer of 2016, the results of which are summarised in~\cite{rosh2018}. The first pulsar science commissioning observations took place in the summer of 2017. We detected single pulses from PSR~B1933+16 and bright giant pulses from PSR~B1937+21 that demonstrated the capability of the instrument to carry out surveys and detect pulsars~\citep{raj2018}. The HI commissioning science results are currently being compiled for a separate publication (Pingel et al.; in prep). 

The PAF consists of 19 dual-polarisation dipoles that are mounted on the dewar producing seven beams on the sky in the pulsar search mode operation. The dewar is cooled using cryo-pumps to provide the user with an unprecedented system temperature over aperture efficiency  ($T_{\rm sys}$/$\eta$). 
From the initial tests, \cite{rosh2018} obtained a  $T_{\rm sys}$/$\eta$= 25~K with an aperture efficiency, $\eta = 0.6$ near 1350 MHz. The lower aperture efficiency compared to the single pixel L-band receiver can be attributed to high ground spillover suppression that is achieved in the formed beams~\citep[see][for more details]{rosh2018, roshi2019}.

The design of the dipole elements mounted on the dewar was optimized for maximum sensitivity over the antenna FoV of angular diameter of $\sim$20' and across a bandwidth of 150~MHz based on prior work~\citep{warnick2011}.
The post-amplification electronics chain is based on an unformatted data digitization technique~\citep{morgan2013} that allowed the entire analog signal path and digitizers to be located directly behind the phased array feed. This enabled a more efficient digitization path and reduced the losses due to transport of analog signals that are prevalent in standard front-end systems. The fiber-optic link transports the digitized samples over 2~km to equipment racks containing five ROACH2\footnote{\url{https://casper.ssl.berkeley.edu/wiki/ROACH}} Field Progammable Gate Array (FPGA) boards, which perform bit and byte alignment, 512 channel polyphase filter bank (PFB), and sideband separation operations. After removing the first and the last 6 band-edge channels, the PFB outputs are sent to a high performance computing (HPC) cluster via 10~GbE links.

The FLAG data processing back-end was developed to produce beamformed spectra as well as raw correlations for HI observations. It consists of five HPC nodes, each equipped with two Nvidia GeForce Titan X Graphical Processing Units (GPUs). The HPCs are connected to a Mellanox 40~GbE switch. The data from the ROACH2 boards are routed through the switch to the different HPCs. Each HPC processes 100 non-contiguous frequency channels. Three basic real-time operation modes are implemented in the HPC cluster: 1. The ``fine'' channelisation mode (Pingel et al.; in prep); 2. Real-time beamformer mode; and 3. A ``coarse'' channelisation or the calibration mode. In the real-time beamformer mode, each GPU runs two processing threads, each forming seven beams over a subband of 25 non-contiguous channels and an effective sampling interval of 130 $\mu$s. The processed subbands are then collated and the resulting 500 channel beamformed spectra for each of the seven beams spanning 150~MHz are written to disk in the VEGAS engineering fits format~\footnote{\url{http://www.gb.nrao.edu/GBT/MC/doc/dataproc/gbtVEGASFits/gbtVEGASFits.pdf}}. A block diagram showing the entire signal chain is shown in Figure.~\ref{fig:pipeline}

\begin{figure}
\centering
\includegraphics[width=\columnwidth]{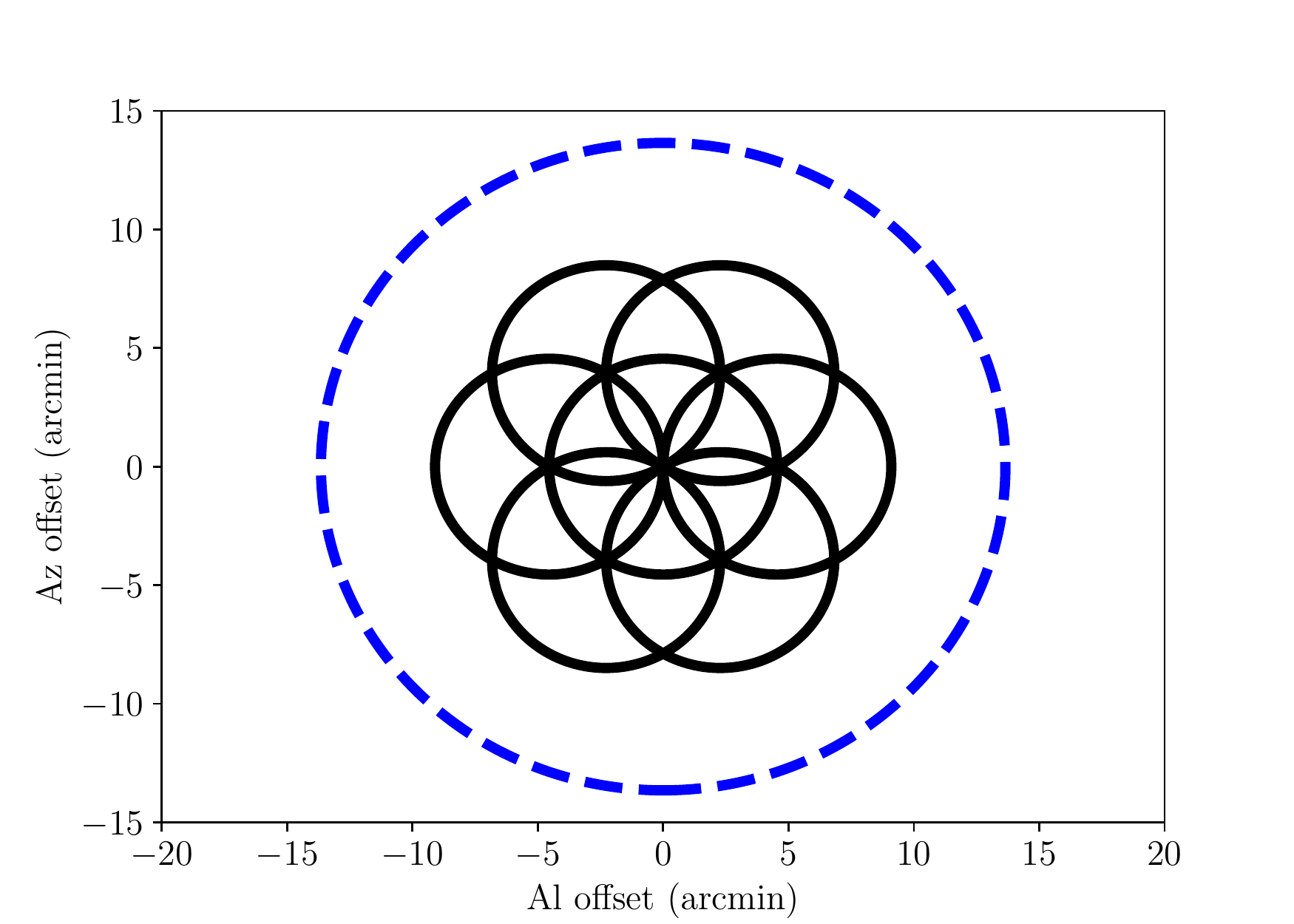}

\caption{ The beam tiling for our survey for single pointing. The black circles enclose the full-width at half maximum (FWHM) of all FLAG beams. The blue dashed circle encloses the area that would be covered if the beams touched at the FWHM.}
\label{fig:beam}
\end{figure}

\begin{figure}
\includegraphics[width=0.5\textwidth]{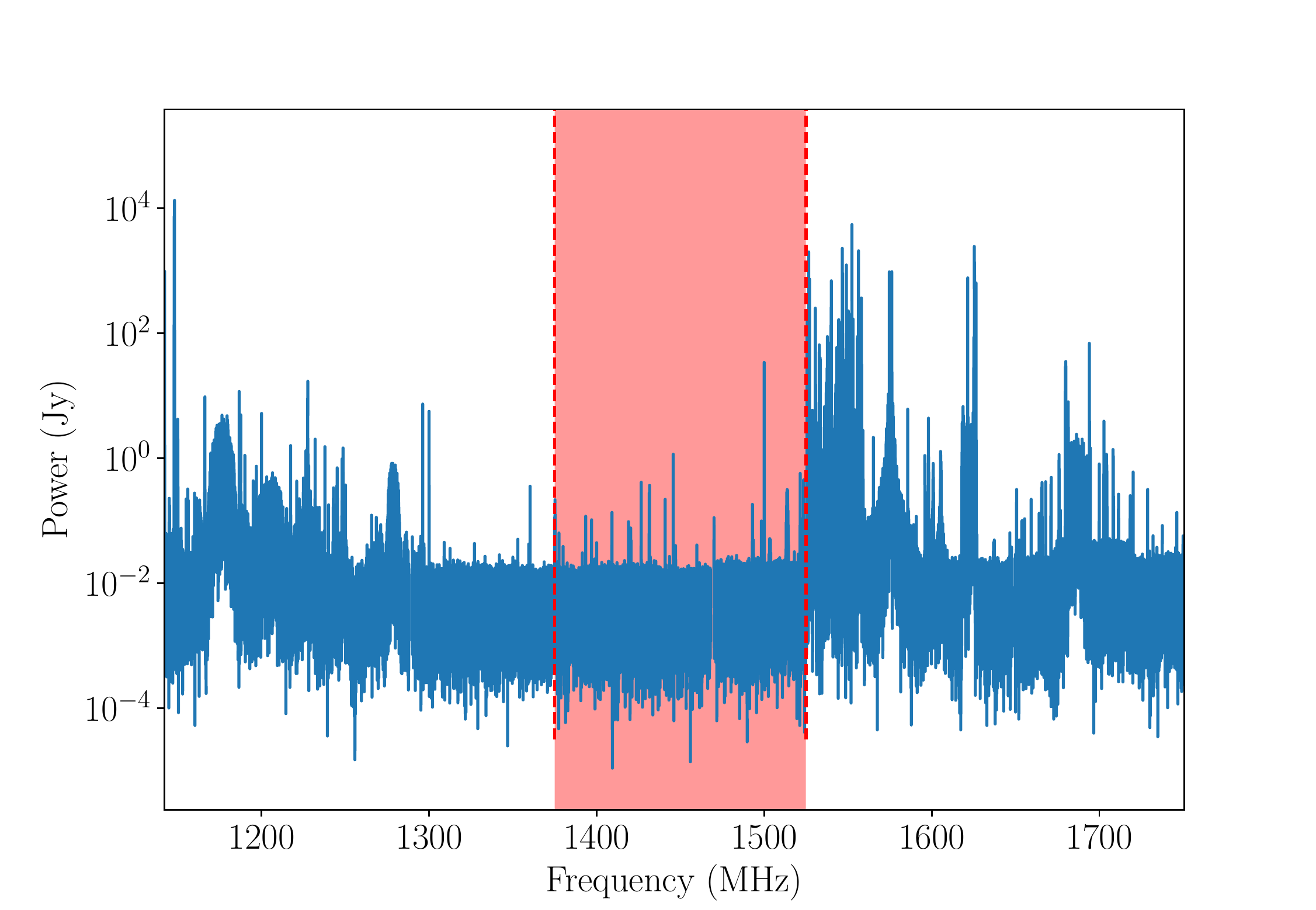}
 \caption{Mean linear polarization power as a function of frequency near 1.4~GHz as measured by the GBT close to the date of our observations. The red shaded region denotes the band selected for the our observations with FLAG.}
\label{fig:rfi}
\end{figure}

\begin{figure*}
\includegraphics[scale=0.34, angle=-90]{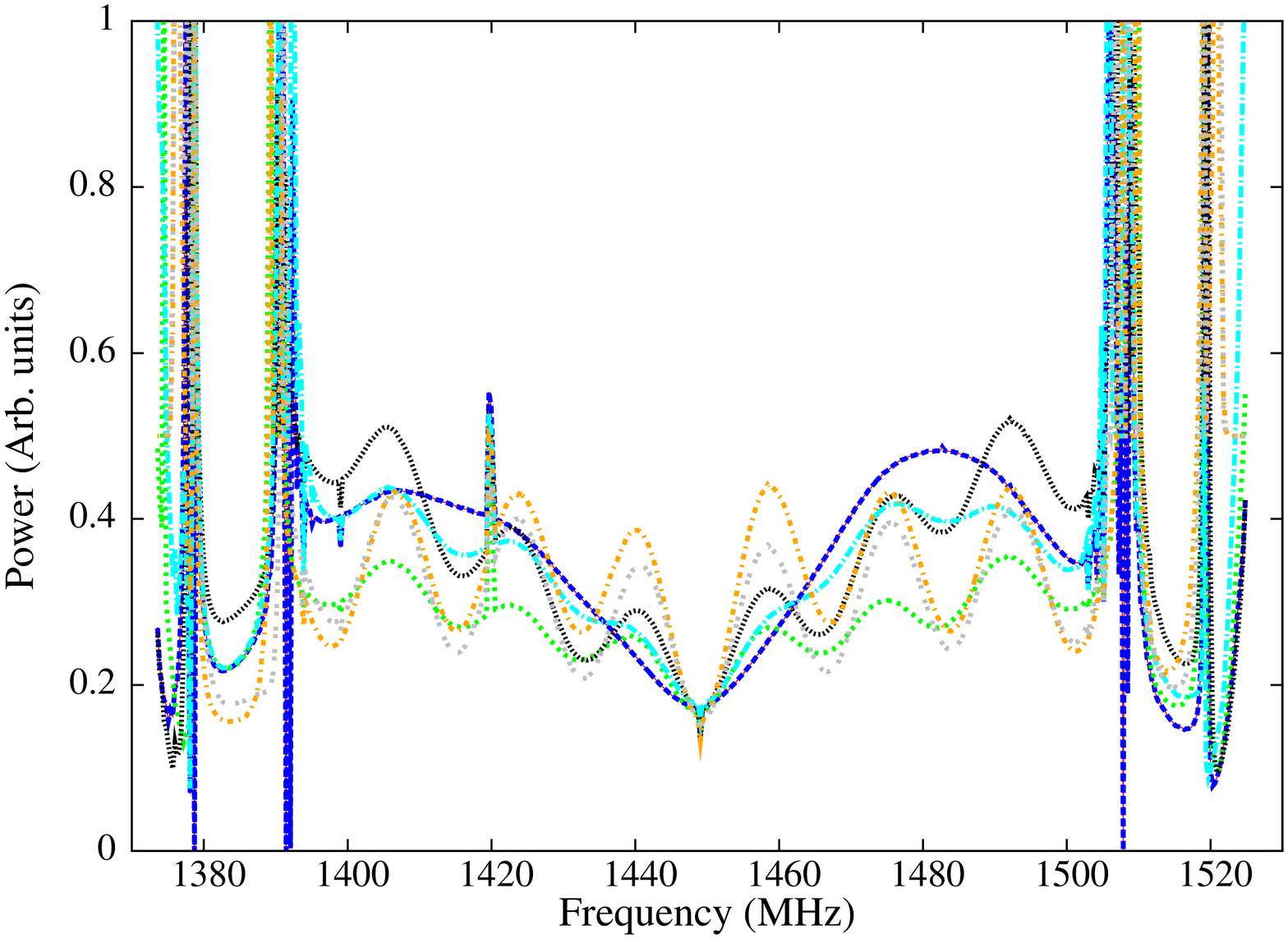}
\includegraphics[scale=0.34, angle=-90]{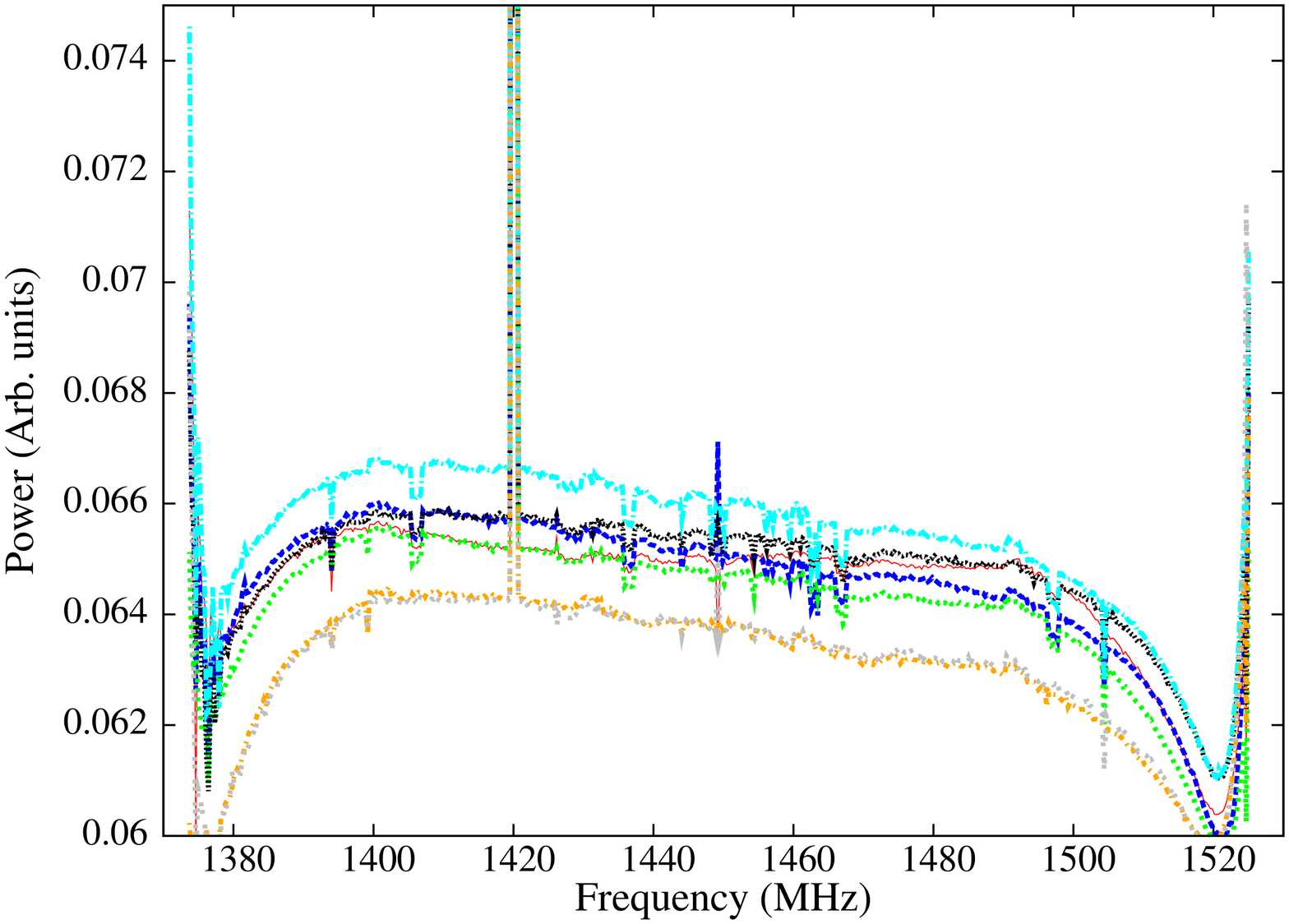}

 \caption{\textbf{Left}: Bandpass plot with central beam pointing at PSR B2011+38 for beam0 (red), beam1 (blue), beam2 (green), beam3 (black), beam4 (cyan), beam5 (orange) and beam6 (grey) for observing session 1 where we faced technical difficulties with the beamformer. \textbf{Right}: Bandpass for all the seven beams for observing session 3. }

\label{fig:bp}
\end{figure*}

\subsection{Pilot Survey}

We conducted a pilot survey for pulsars and FRBs using FLAG. We observed high Galactic latitudes with telescope pointings spanning Galactic latitudes 75$^\circ$ to 78$^\circ$ and Galactic longitudes 30$^\circ$ to 33$^\circ$ split between three observing sessions: 28th January 2018, 1st February 2018 and 4th February 2018. Though the expected density of pulsars off the Galactic plane is low, we chose this region of the sky for two main reasons. Firstly, this region of the sky has not been extensively searched for pulsars by previous surveys, providing us with a good opportunity to discover pulsars. Secondly, because the survey region is off the Galactic plane, it improves our chances of finding faint MSPs due low scattering along the line of sight~\citep{bh04} and we can expect to find MSPs off the Galactic plane as they are an older population~\citep{le13}. 

We used an integration time of 300~s per pointing. Before each observing session, a flux calibrator 3C295 was observed at the boresight of all seven beams with two reference off-source pointings that were 2 degrees away in cross-elevation from the top most and the bottom most beam in elevation to obtain the beamformer weights using the maximum signal-to-noise (S/N) ratio methodology~\citep{elmer2012} for the seven beams.

\begin{figure*}
\includegraphics[width=0.6\textwidth]{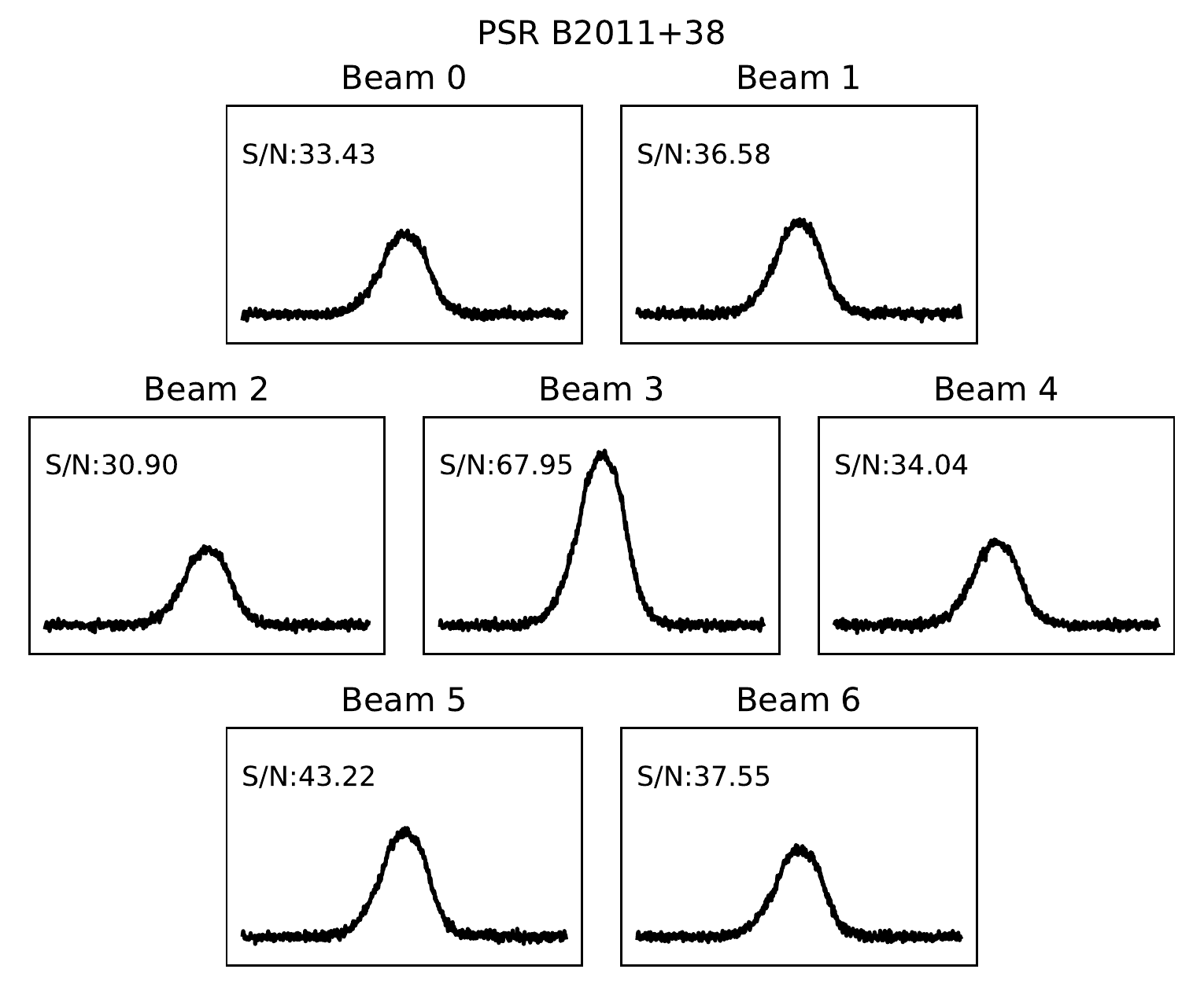}
 \caption{Observation of PSR~B2011+38 in the central beam with Nyquist separation of other beams. The reduction in S/N of the pulsar by a factor of two in surrounding beams is evident.}
\label{fig:7beam}
\end{figure*}

The expected sky coverage from the survey was $\sim$10~deg$^{2}$, but beams were located at Nyquist separation to optimize an earlier set of HI observations. As a result, our effective sky coverage was one third smaller than the maximum possible sky coverage (see Fig.~\ref{fig:beam}). During each observing session, we observed a test pulsar, namely PSR~B2011+28 in all the beams to ensure that the beamformer weight calibration is correct and that the weights are applied correctly in the real-time beamformer system.

 Since the beams were placed at the Nyquist separation, as shown in Figure~\ref{fig:beam}, a pulsar observed with the central beam was also detected in each of the other six beams with a factor of two lower S/N compared to that obtained in the central beam. Although this test verified the sensitivity of each of our beams, it did make it harder to differentiate between true astrophysical events and radio frequency interference (RFI).
Due to the prevalence of strong RFI at the GBT in this frequency band, we decided to shift the Local Oscillator (LO) such that the centre frequency of the 150~MHz band would be in the clean part of the frequency range. Figure~\ref{fig:rfi} shows the average power in every channel in our band close to the time of our observations. One can clearly see the RFI beyond 1.5~GHz due to telecommunication satellites.

We experienced some technical issues with the beamformer during the first session of observations which spanned nearly half of the total survey time. During the calibration grid scans for generating beamformer weights, a wrong value for the quantization gain caused the generation of incorrect weights that significantly reduced the sensitivity of the formed beam and also introduced some artifacts in the signal.  Figure~\ref{fig:bp} shows the bandpass from one such observation where the artifacts due to incorrect calibration are clearly visible in comparison to the expected bandpass during session 3 which was mostly flat with a strong detection of the HI line. Hence, we decided to not use the first session of observations for the search. The technical difficulties encountered during the first observing session reduced the total searched area and the total time on the sky by 50$\%$. 

For each pointing, the data for all seven beams were recorded using the FLAG beamformer (see Fig.~1). Then, for each time sample, each frequency subband consisting of 25 non-contiguous channels was  manipulated and merged to a contiguous 500 channel spectrum. Each of these consecutive spectra were collated and converted into a filterbank with the necessary metadata obtained from the telescope monitoring system. All of these steps were performed using our custom built FLAG beamformer pulsar processing software\footnote{\url{https://github.com/krajwade/FLAG-Beamformer-pulsar}}. 

Standard pulsar data processing software like SIGPROC~\footnote{\url{http://sigproc.sourceforge.net}} and PRESTO~\citep{ran2011} expect each spectrum to be inverted (high to low frequency order) hence the bandpasses for each beam and each time sample were flipped before being written out to disk. The resulting ``filterbank'' files were used for further data analysis.

\begin{table}
	\centering
	\caption{Survey Parameters for the FLAG pilot survey}
	\label{tab:survpar}
	\begin{tabular}{lr} 
		\hline
		Number of Beams & 7\\
		Number of Polarizations ($n_{p}$) & 2\\
		Centre Frequency (MHz) & 1440\\
        Frequency channels & 500\\
        Bandwidth ($\nu$, MHz) & 150 \\
        System Temperature (T$_{\rm sys}$, K) & 30\\
        Galactic Longitude range & 75$^{\circ}$ --  78$^{\circ}$ \\
        Galactic Latitude range & 30$^{\circ}$-- 33$^{\circ}$ \\
        RA range (J2000) & 17h 31m 59.1s -- 17h 50m 54s \\
        Declination range (J2000) &  48$^{\circ}$ 27m 22.4s -- 50$^{\circ}$ 57m 54.1s\\
        Sampling interval & 130~$\mu$s\\
        Integration time ($\tau_{\rm int}$, s) & 300 \\
        Telescope Gain (K~Jy$^{-1}$) & 1.7\\
		\hline
	\end{tabular}
\end{table}

\subsection{Survey Sensitivity}

\begin{figure*}
   \includegraphics[scale=0.27]{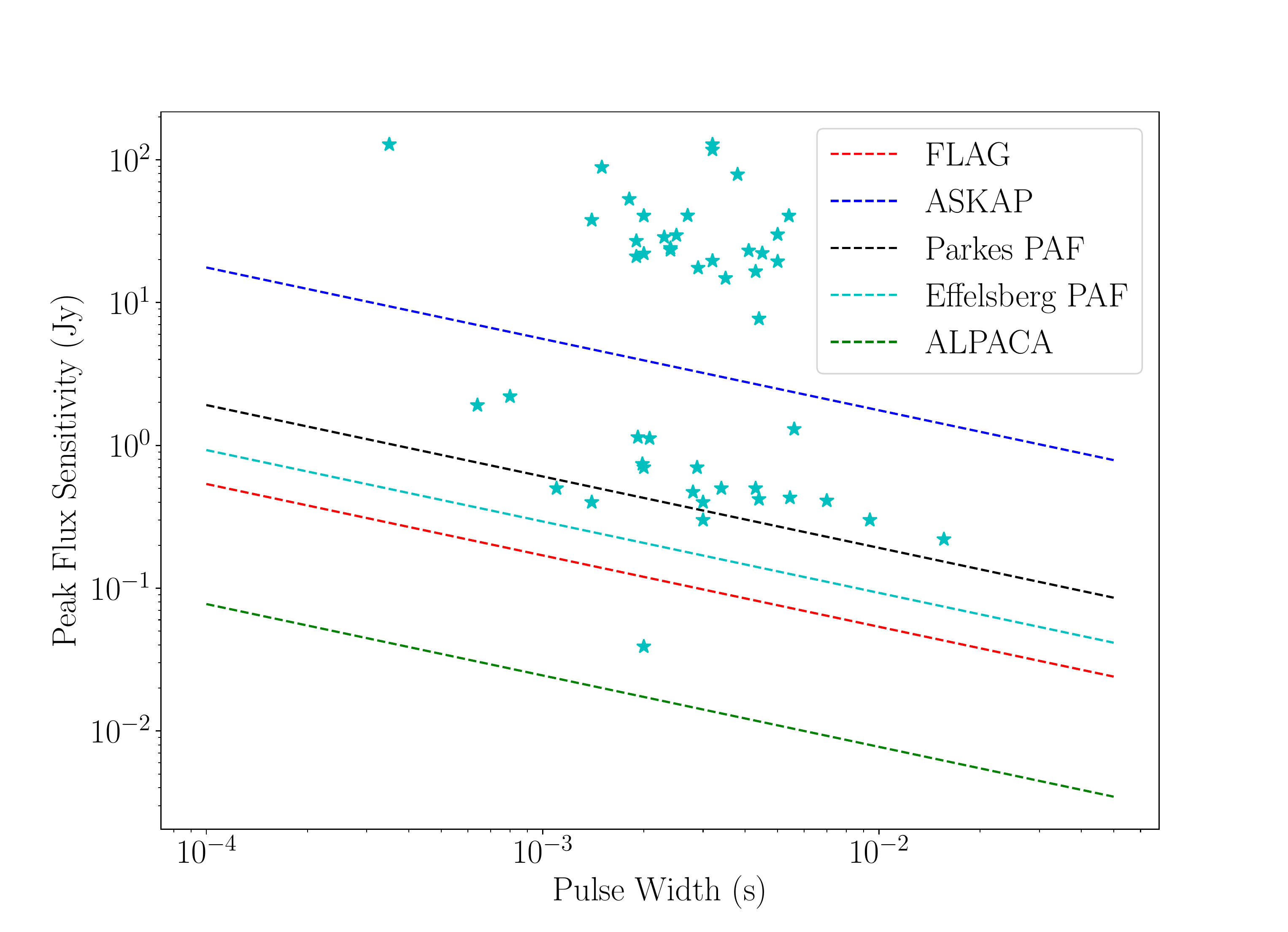}
    \includegraphics[scale=0.275]{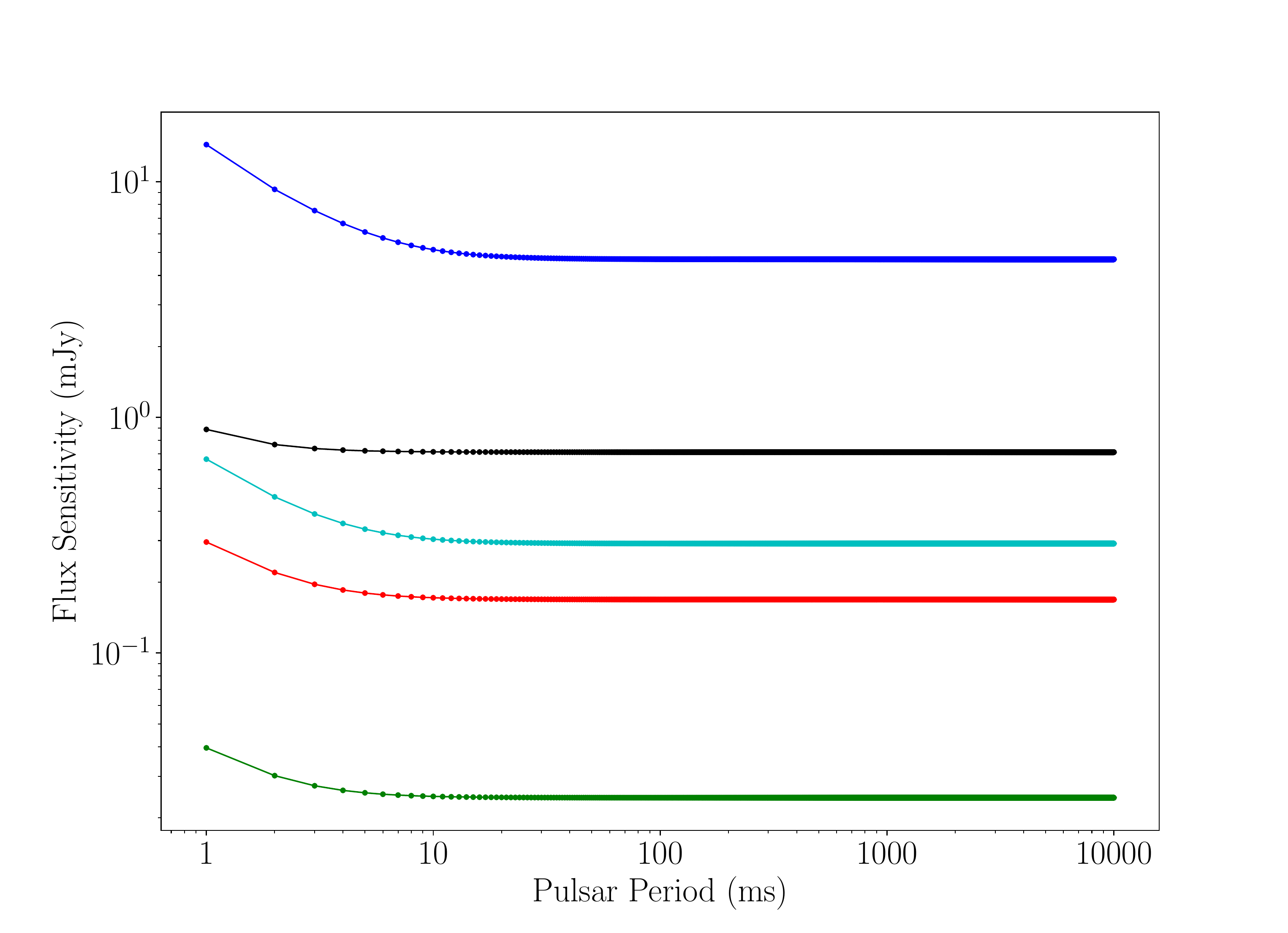}
\caption{\textbf{Left}: Peak flux sensitivity of PAF surveys to bright single pulses as a function of pulse width along with detected FRBs (cyan stars) taken from the FRB catalog~\citep{pet16}. We assumed a scattering width of 8~ms for the limit calculation based on values from~\citet{cor16}. \textbf{Right}: Flux limit of surveys with different PAFs as a function of pulsar period. The maximum DM for the search is 100~pc~cm$^{-3}$. The sensitivities for each of the PAFs were calculated based on the parameters given in Table.~\ref{tab:lim}}.
\label{fig:survlim}
\end{figure*}

Based on our survey parameters given in Table~\ref{tab:survpar}, we can calculate the survey sensitivity. For a minimum detection threshold (S/N$_{\rm min}$), the limiting flux of a survey is
\begin{equation}
S_{\rm min} = \frac{\rm S/N_{\rm min}~T_{\rm sys}~\beta } {G\sqrt{n_{p}~\Delta \nu~\tau_{\rm int}}} \sqrt{\frac{W_{\rm eff}}{P-W_{\rm eff}}},
\label{eq:perflux}
\end{equation}
where,
\begin{equation}
W_{\rm eff} = \sqrt{ W_{\rm int}^{2} + W_{\rm DM}^{2} + W_{\rm \tau}^{2} + W_{\Delta t}^{2}}.
\end{equation}
Here, $T_{\rm sys}$ is the system temperature (sum of the receiver temperature and sky temperature), $\beta$ is the degradation factor due to digitisation, $G$ is the instrumental gain, $n_{p}$ is the number of polarisations to be summed, $\Delta \nu$ is the bandwidth of the backend and $\tau_{\rm int}$ is the integration time. The effective width of the pulse, $W_{\rm eff}$, is the quadrature sum of the intrinsic width, W$_{\rm int}$, intra-channel smearing due to dispersion, $W_{\rm DM}$, smearing due to interstellar scattering, $W_{\rm \tau}$ and smearing due to finite sampling interval, $W_{\rm \Delta t}$.  Since the first session was not used for any searches, we compute the flux limit for the last two sessions using Eq.~\ref{eq:perflux}.  The contribution to the system temperature due to the Galaxy is minimal ($\sim$1~K) as we were observing off the Galactic plane. Moreover, the effective width of the pulse will not be affected drastically by the DM smearing and scattering since the maximum line of sight DM is 47~cm$^{-3}$~pc based on the NE2001 electron density model~\citep{co02}. This makes the region of the sky advantageous for searching faint MSPs as scattering heavily affects their detection threshold. The right panel of Figure~\ref{fig:survlim} shows the flux limits of FLAG in the aforementioned region of the Galaxy with the expected sensitivity of the contemporary PAFs at other telescopes as a function of pulse period. One can clearly see that FLAG is more sensitive compared to most other PAFs by an order of magnitude.

For the FRB searches, we computed the fluence threshold of our search. For a given intrinsic width of the FRB, $W_{\rm int}$, the peak flux sensitivity limit,
\begin{equation}
S_{\rm lim} = \frac{\rm S/N_{\rm lim} T_{\rm sys}}{ G~W_{\rm int}} \sqrt{\frac{W_{\rm eff}}{n_{p} \Delta \nu}},
\label{eq:frbflux}
\end{equation}
where the terms have the same meaning as in the previous equation.
The left panel of Figure~\ref{fig:survlim} shows the sensitivity limit to detect FRBs for various PAFs which also demonstrates the high sensitivity of FLAG. One must note that the plot does not take into account the FoV of the PAFs. Since the FoV of other PAFs is larger than the FoV of the FLAG by  a factor of 5--15 and since we covered a  small portion of the sky with a small survey time, the prospects of finding a FRB in a small survey area with the FLAG are remote. On the other hand, ASKAP, though less sensitive compared to  FLAG, has already discovered more than 20 FRBs owing to a larger FoV and extensive time on sky (Shannon et al. 2018). We note that this limit also applies to bright single pulses emitted by pulsars and rotating radio transients \citep[RRATS;][]{mc06}.

\section{Survey Processing Pipeline}
We processed the data from two of the three observing sessions through our single pulse search and periodicity search pipelines. Below we describe each pipeline along with the validation of the same using a test pulsar observation.

\subsection{Single Pulse Search}
To search for FRBs we used a GPU based dedispersion pipeline {\sc heimdall}\footnote{\url{https://sourceforge.net/projects/heimdall-astro}}. The data were de-dispersed for 874 trial DMs from 20 to 10000~pc~cm$^{-3}$. The trial DMs were chosen such the S/N of a pulse of 40~ms would result in a S/N drop of 25\% at the next DM trial. Then, the dedispersed time-series were convolved with nine box car functions of widths 130~$\mu$s to 66.56~ms, with increasing box car widths of consecutive power of 2 number of time samples, and the resulting timeseries were searched for bright pulses with S/N~$> 6$. The candidates above our S/N threshold were then clustered together in DM/arrival time/width parameter space. The clustering results in a significant reduction in the number of candidates that need to be inspected. The resulting candidates with S/N~$>8$ and with more than five members in the cluster were then chosen for manual inspection.

\begin{figure}
\centering
\includegraphics[width=\columnwidth]{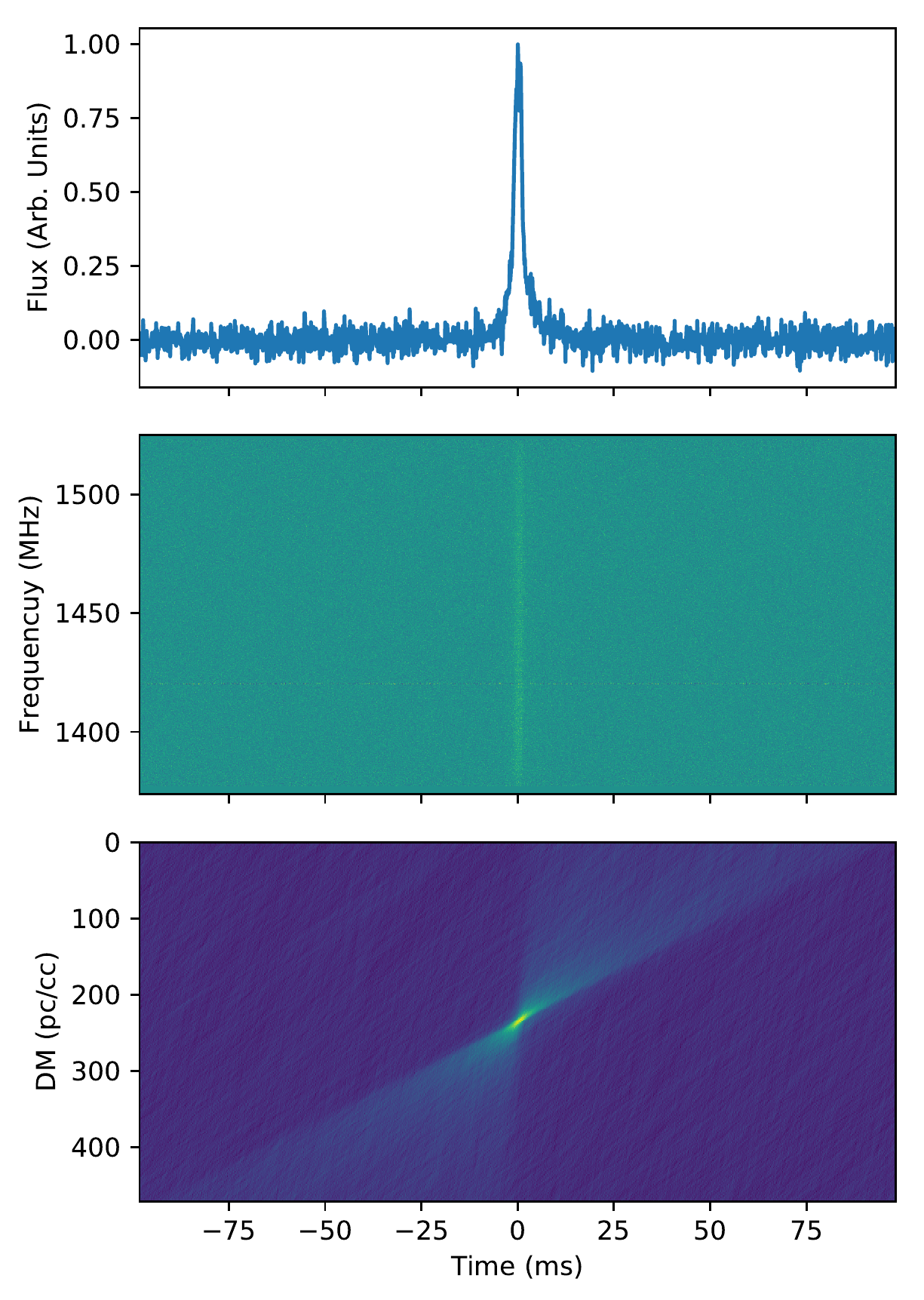}
\caption{A 100$\sigma$ single pulse from PSR~B2011+38 detected by our pipeline. The top plot shows the pulse profile, the middle plot shows the dedispersed spectra, and the bottom plot shows the DM--time plot.}
\label{fig:B2011_sp}
\end{figure}
Due to our Nyquist sampled beams, the candidates from different beams could not be coincidence filtered (i.e. checked whether the same burst is seen in all the beams to differentiate between astrophysical signals and RFI). Figure~\ref{fig:B2011_sp} shows a 100$\sigma$ single pulse of PSR~B2011+38 detected through the above-detailed pipeline. For FLAG's sensitivity the expected average single pulse S/N for this pulsar is $\sim$5 therefore we expected about 30$\%$ of the single pulses to be detectable assuming a log-normal distribution of single pulse energies~\citep{burke2012}. We detected 905 single pulses from PSR~B2011+38 above a S/N of 8 in our 11-minute test observations which amounts to 31$\%$ of the total pulses. The search resulted in a total 19756 candidates to be inspected visually that turned out to be mostly broadband terrestrial RFI. However, no new transients of convincing astrophysical origin were detected. 

\subsection{Periodicity Search}
\begin{figure*}
\centering
\includegraphics[trim = 300 0 0 0, clip, width=\textwidth]{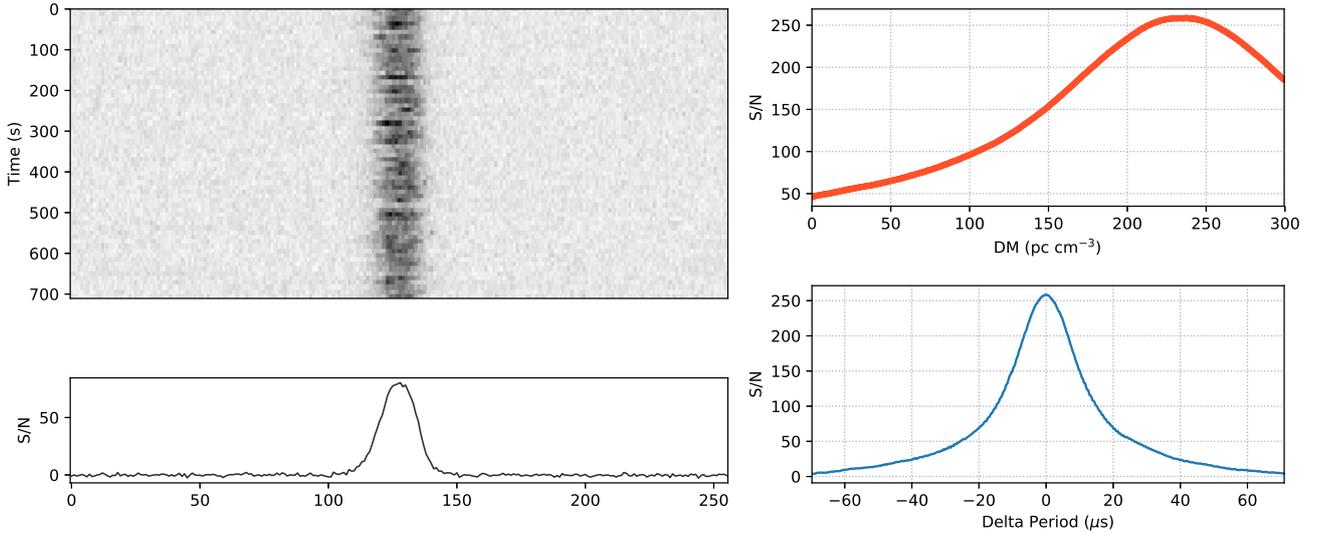}
\caption{A detection of PSR~B2011+38 using the FFA pipeline. The grey-scale plot shows the folded time series of the candidate along with the folded profile at the bottom. Top right panel shows S/N as a function of DM and bottom right panel shows the S/N as a function of period offset.}
\label{fig:B2011_ffa}
\end{figure*}

To detect periodic signals in our dataset, we de-dispersed the generated filterbank data over a range of trial DMs using standard PRESTO software pipeline. Based on the maximum DM along the line of sight of 47~cm$^{-3}$~pc, we used a DM range from 1--100~cm$^{-3}$~pc in steps of 0.5 DM units for our periodicity searches. Then, to search for pulsars, we took a two pronged approach. Recent advances in accelerated computing have brought the Fast Folding Algorithm~\citep{st69} to the fore~\citep{pa2018}. While standard Fourier techniques have been used for searching for pulsars since their discovery~\citep{lo04}, they can be less sensitive due to the loss of power during incoherent harmonic summing~\citep[see][for more details]{raj2018b}. The FFA, on the other hand, is a time domain search technique that folds the time series at a set of trial periods to avoid the loss of power and will essentially be more effective at finding pulsars.  We must note that the FFA does not scale linearly in terms of computation cost as we go to shorter periods and it can get quite computationally expensive. Hence, we decided to split the periodicity search based on the searched period range since FFT searches are less sensitive for longer periods due to effects of red noise in the data, we used the FFA algorithm for longer periodicities. To search for millisecond pulsars (MSPs), we used standard Fourier algorithms since the FFA becomes computationally unfeasible for extremely small periods~\citep{cam17}. Figure~\ref{fig:B2011_ffa} shows the detection of PSR B2011+28 in the FFA search. The implementation of the FFA algorithm will be presented in another paper (Morello et al., in prep).

For the Fourier searches, we first used standard RFI mitigation techniques in PRESTO to remove narrow-band frequency domain RFI. Then, each dataset was converted to time series over the range of DM trials mentioned above. For each DM trial, we ran the \texttt{accelsearch} that performs frequency domain periodicity search over the time series. For the acceleration search, we looked for signals drifting by $\pm$200/N bins in the Fourier domain, where N is the largest harmonic at which a signal is detected above a threshold of 5.0. We summed to a maximum of 16 harmonics for the search. Then, the detected candidates were saved for visual inspection. For the FFA search, the de-dispersed timeseries were further used for time-domain folding. We used the multi-threaded FFA pipeline to process multiple DM trials simultaneously to improve the speed of the FFA search. We searched for periodicities from 300~ms to 10 s using the FFA, while for shorter periods, we used Fourier techniques.

\section{Results and Limits}
We searched for pulsars and FRBs in session 2 and session 3 spanning 5 hours of data in total. We did not find any FRBs, pulsars or RRATs in the searched data.  Based on our predictions, we expected to find 1--2 pulsars in the full survey region. The lack of new pulsar or FRB discoveries is not unexpected (See Section 5). The non-detection of FRBs helps in placing upper limits on the all-sky FRB rate above the detection fluence. Most simply, we calculate the upper limit to the rate as the rate at which our survey should have found 1 FRB based on our sky coverage and total integration time. Based on that, we find R = 5.9$\times$10$^{6}$ FRBs per sky per day. 

Recently with the discovery of FRBs from ASKAP, \citet{sh2018} has reported an event rate ($R_{\mathrm{ASKAP}}$) of $37 \pm 8 $ per day per sky above a fluence limit of 26~Jy~ms. For our survey, with a bandwidth ($\Delta \nu$) of 150~MHz, pulse width ($W_{\rm eff}$) of 1.26~ms a signal to noise (S/N) threshold of 10, our fluence limit is given by,
\begin{equation}
   \mathscr{F}_{\mathrm{FLAG}} = S/N_{\rm lim} \frac{T_{\mathrm{sys}}}{G}\sqrt{\frac{W_{\rm eff}}{n_p \Delta \nu}}.
   \label{eq:fluence}
\end{equation}
Since FRBs are typically detected away from the boresight of the beam, we estimated a conservative limit on the fluence by computing it at the FWHM of the beam. Using the fluence limit from the above equation, we can scale the ASKAP rate to our conservative fluence threshold using Eq.~12 from~\cite{ch17}. Thus,
\begin{equation}
    R (> \mathscr{F}) = R_{\mathrm{ASKAP}} \Bigg( \frac{\mathscr{F}}{26~\mathrm{Jy~ms}} \Bigg)^{\gamma},
\end{equation}
where $\gamma$ is the source count index. Assuming a $\gamma$ = -1.5 for a Euclidean source count distribution, we obtain  R( $>0.36$~Jy~ms) = $(2.2 \pm 0.5) \times 10^{4}$ FRBs per sky per day. We must note that our survey was more sensitive compared to the ASKAP survey and there are large uncertainties involved in the intrinsic slope of the source count distribution of FRBs~\citep{mac2018}. Recently,~\citet{ja19} have shown that the slope of the source count distribution flattens between the ASKAP and Parkes FRB samples which can lead to large uncertainties when we scale the ASKAP rates to the fluence limit of our survey. To get a better handle on the upper limit, we assume that FRBs follow a Poisson-Point process. Then, we use techniques presented in~\cite{ghe1986} to compute the 90$\%$ confidence level upper limit on the number of detected events. For a given rate, $R$, survey duration, $T$, and sky coverage, $\Omega$, the 90$\%$ confidence level upper limit on the number of events,
\begin{equation}
    N_{u} = -\ln(0.1) = 2.303.
\end{equation}
Then the upper limit on the rate can be estimated by dividing $N_{u}$ by product of $T$ and $\Omega$. Using the this method, we report our 90$\%$ confidence level upper limit on the rate to be 1.3$\times$10$^{6}$ FRBs per day per sky. We note that we used the reduced sky coverage and the reduced survey time for all the rate calculations in this section.

\begin{figure}
\centering
\includegraphics[width=\columnwidth]{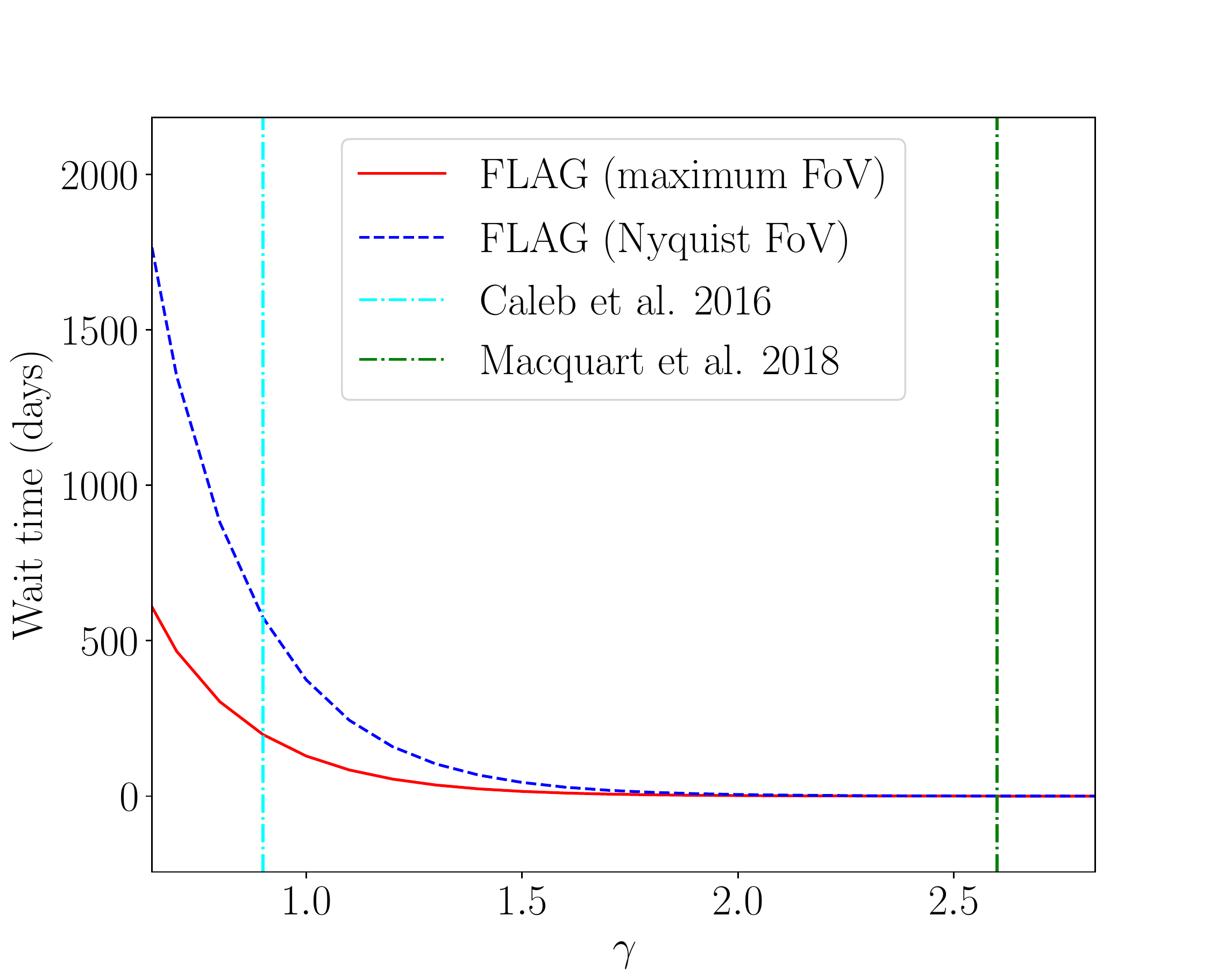}
\caption{FRB detection wait time in days as a function of the slope of the source count distribution of FRBs. The plots are shown for two different FoVs of the FLAG PAF (see text for more details). The vertical lines correspond to limits on the slope from the literature.}
\label{fig:frb}
\end{figure}

Since the rate of observed events depends on the effective FoV of the PAF, we have computed the predicted wait times to observe an FRB for FLAG for various FoVs. To do this, we first compute the maximum fluence limit of FLAG by computing it at FWHM of the beam using Eq.~\ref{eq:fluence}. Then, we can scale the ASKAP rate to the this fluence limit to obtain a conservative rate for FLAG that takes into account the flux degradation of the detected FRB due to the offset from the boresight of the beam. We also note that the calculations for the FWHM were done assuming a Gaussian beam. Then, the time it takes to observe 1 FRB,
\begin{equation}
T = \left(R \Omega \right)^{-1},
\end{equation}
where $\Omega$ is the sky coverage of the FLAG PAF and $R$ is the rate. We computed $T$ for a range of $\gamma$s for the maximum FoV of FLAG and our observations. Those results are shown in Figure.~\ref{fig:frb}. Based on our results, it is clear that for a flat source count distribution, a survey with maximum FoV would detect an FRB a factor of 2 faster compared to Nyquist sampled FoV while the detection wait times become similar for steeper slopes. 

\section{Discussion}

\begin{table*}
\begin{minipage}{\textwidth}
\begin{tabular}{lc lc lc lc lc lc}
\hline
Telescope(PAF) & Central Frequency & Bandwidth & System Temperature & Gain & FoV & N$_{b}$ &Reference \\
\hline
      & (GHz) & (MHz) & K & K Jy$^{-1}$ & deg$^{2}$ & &\ \\
\hline
  ALPACA &  1.374 & 300 & 30 & 10.0  & 0.1 & 40& \protect~\cite{cortes2016}\\
 Effelsberg PAF & 1.367 & 303 & 65 & 1.8 & 0.65 & 36& \protect \cite{mal2018}\\
 Parkes PAF &  1.374 & 303 & 65 & 0.6 & 1.53 & 36& \protect \cite{deng2017}\\
 FLAG & 1.440 & 150 & 30 & 1.7 &  0.1 & 7 &This paper\\
GBT L-band receiver & 1.4 & 640 & 30 & 2.0 & 0.0015 & 1& GBT Proposer's Guide~\footnote{\url{http://www.gb.nrao.edu/scienceDocs/GBTpg.pdf}} \\
\hline
\end{tabular}
\caption{Assumed system parameters for different PAFs used in this simulation. Here, FoV is the instantaneous field of view and N$_{b}$ is the number of beams on the sky.}
\label{tab:lim}
\end{minipage}
\end{table*}

\begin{figure*}
   \includegraphics[width=.49\linewidth]{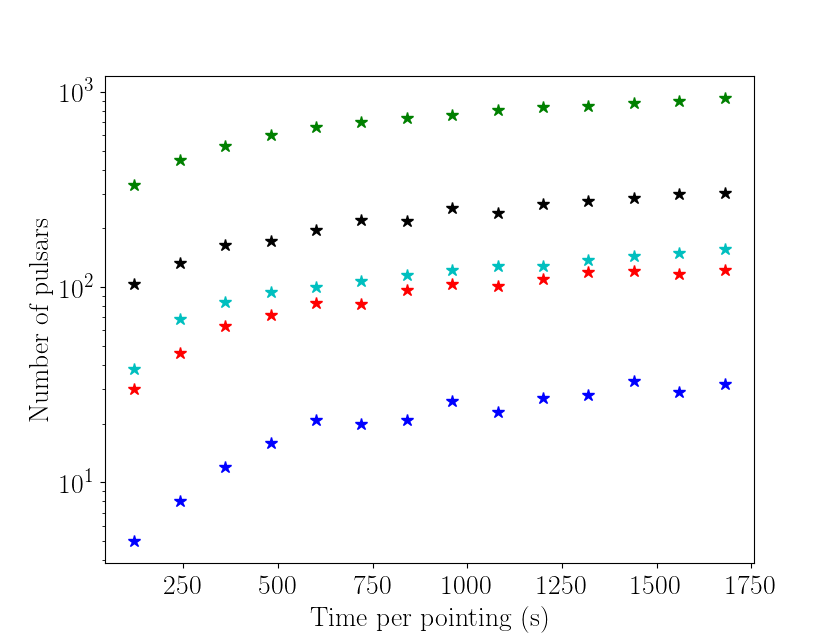}
    \includegraphics[width=.49\linewidth]{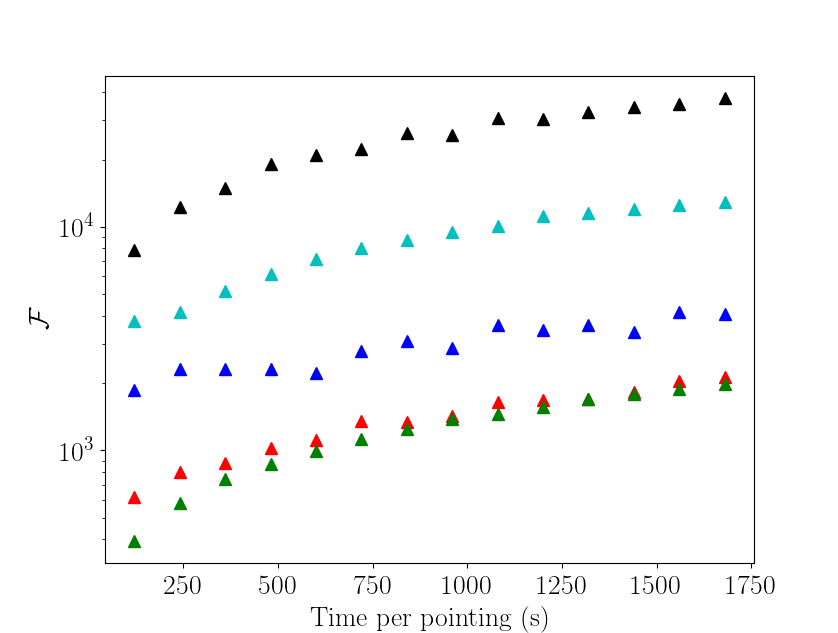}
\caption{\textbf{Left}: number of pulsars detected in a putative Galactic plane survey using Parkes PAF (blue stars), Effelsberg PAF (red stars), FLAG (cyan stars), GBT L-Band single pixel receiver (black stars) and ALPACA (green stars) as a function of integration time per pointing. \textbf{Right}: survey metric for the same surveys as a function of integration time per pointing.}
\label{fig:PAF_sim}
\end{figure*}

PAF technology has greatly advanced over the last decade. The availability of PAF receivers on single dish telescopes implies that one can carry out all sky surveys at the same sensitivity in a much shorter time. PAF receivers are being developed and used at most single dish telescopes around the world. The PAF receiver at the Parkes~\footnote{the same PAF has now been moved to Effelsberg  (M. Keith, private communication)} and Effelsberg telescopes have been recently used for carrying out test observations of pulsars and RRATs ~\citep{deng2017,mal2018}. What makes the FLAG receiver stand out compared to these PAFs is that it is a cryo-cooled receiver that gives an unprecedented sensitivity in carrying out searches for pulsars. However, we note that the PAF at Parkes or at Effelsberg has a larger field of view and a larger bandwidth. Hence, these PAFs can spend more time on a pointing to obtain the same sensitivity as FLAG and still finish the survey of a given region in half the time. To compare the efficacy of a pulsar survey with different PAFs, we ran simulations for expected number pulsar of discoveries using different PAFs if they were to conduct a Galactic plane Survey. To do that, we used PsrPopPy\footnote{\url{https://github.com/samb8s/PsrPopPy}} ~\citep{ba14}, a pulsar population synthesis suite, to generate a synthetic population of pulsars in the Galaxy that mimics the observed characteristics of known pulsars~\citep[see][for more details]{ba14}. Using this software suite, we generated a snap-shot population of pulsars in the  Galaxy based on some basic assumptions on the probability distribution functions of various parameters like spin period, surface magnetic field strength, scattering timescales and spatial distribution in the Galaxy (For a complete set of assumptions, see Bates et al.~2014). To make sure that our simulated population represents the true pulsar population in the Galaxy, we calibrated our simulated population to the Parkes multi-beam pulsar survey~\citep{ma01a} such that the number of discoveries in a simulated Parkes multi-beam survey over our population would result in a similar number of discoveries as compared to the actual survey. Using this synthetic population, we performed a  simulated Galactic plane survey for the three PAFs namely, Effelsberg PAF, Parkes PAF and the ALPACA receiver. To do this, we ran the simulated search over Galactic longitudes from --30$^\circ$ to 30$^\circ$ and latitudes of --5$^\circ$to 5$^\circ$ as this region is visible to all of these PAFs. We calculated the sensitivity and the FoV of each PAF based on the parameters that are summarized in Table~\ref{tab:lim}. Then, for every simulated pulsar within the survey region, we counted how many of them were within the sensitivity threshold of the survey after taking into account effects like multi-path scattering, dispersion smearing and flux degradation due to off-axis location of the pulsar in the beam. We repeated this analysis for every PAF for different integration times. To measure the ability of the PAF receiver to find pulsars, we need to take into account not only the sensitivity of the receiver but also the FoV that matters when there is limited time available to cover large portions of the sky.  Hence, for a given pulsar survey, the measure of success can be defined by the time it takes to find one pulsar,
\begin{equation}
    \mathcal{F} = \frac{N_{\rm pointings} \times \tau_{\rm pointing}}{N_{\rm psr}} ,
\end{equation}
where $N_{\rm pointings}$ is the total number of pointings, $N_{\rm psr}$ is the total number of pulsars discovered in the survey and $\tau_{\rm pointing}$ is the total time per pointing in the survey. Figure~\ref{fig:PAF_sim} shows the results for our simulations. It can be seen clearly that a pulsar survey of the Galactic plane with the FLAG would be able to discover more pulsars compared to the PAFs on Parkes and Effelsberg radio telescope if we assume that each survey spends the same amount of time per pointing. This means that FLAG will be able to go deeper than any other PAF to find fainter pulsars. Since the FoVs of other PAFs are larger by factors of 5--15, they would perform better as survey instruments as seen in the right panel of Figure~\ref{fig:PAF_sim}. It means that though FLAG is more sensitive compared to other PAFs, their larger FoV more than makes up for  the lower sensitivity i.e., they can observe a pointing for longer to achieve the same sensitivity and still finish the entire survey faster. In spite of this shortcoming, FLAG will perform faster surveys compared to the current single pixel L-Band feed at the GBT by a factor of 2--3. We also show results from ALPACA, which will form 40 beams on the sky totalling a FoV of 0.1~deg$^2$. We see that it will be the most sensitive PAF to date and will discover pulsars quicker than all PAFs by a factor of 2--10 for short duration pointings. For longer dwell times,  ALPACA is expected to be comparable to the Effelsberg PAF. Overall, our results show that PAFs are extremely useful survey instruments and will be the primary survey instruments for single dish telescopes in the near future and that FLAG will be the premier survey instrument for pulsars and FRBs at the GBT.

Moreover, PAFs  have the added advantage of larger FoV to search for FRBs. In our pilot survey, we did not detect any FRBs; this can be attributed to the small sky coverage of the survey in spite of the greater sensitivity. We also computed the predicted wait times to detect FRBs with different FoVs of the receiver. We can clearly see that if the slope of the source count distribution is flatter then FoV becomes an important factor and FLAG would have to survey the sky for $\gtrsim$ 100 days to detect an FRB. On the other hand, for steeper slopes, the sensitivity becomes more significant and the wait times can be as small as $\sim$2 days for different FoVs of the FLAG.

The use of PAFs for FRB surveys makes single dish telescopes more relevant for FRB science in comparison with interferometers that are being extensively used for finding and localising FRBs~\citep{sh2018, ban2019}. The ability to form multiple beams in different locations of the sky gives PAFs the capability to localise FRBs with much better accuracy than what would be achieved with single pixel receivers. The ability of forming beams flexibly gives PAFs an advantage over multi-reciever systems like Arecibo L-band Feed Array (ALFA) and the Parkes multi-beam receiver as they can form overlapping beams to enable better localisation. Moreover, one can use techniques developed for interferometers with complex beam morphologies to localize the position of the source to a much higher accuracy compared to what would be achieved from detections in multiple beams.~\citep[][and the references therein]{ob2015}. This means that future pulsar and FRB surveys with single dish telescopes will spend less time in localising the burst compared to single-pixel receivers.

\section{Conclusion}
Here, we have presented results from the pilot survey for pulsars and fast radio bursts with the newly commissioned FLAG receiver at the GBT. We covered approximately 10 deg$^{2}$ of sky in a span of about 10 hours. One of the three observing sessions had technical issues that rendered the data from that session unsuitable for the search. Though we did not find any pulsars or FRBs, we were able to demonstrate the capabilities of the instrument. Using test pulsar observations during the survey, we showed that we could detect bright single pulses in our single pulse search pipeline as well as detect pulsars in the FFA search pipeline to look for periodic astrophysical signals. Based on our non-detection, we report a 90$\%$ confidence level upper limit on the FRB rate of 1.3 $\times 10^{6}$ FRBs per sky per day for a fluence limit of 0.36~Jy~ms. We also show that PAF receivers are better suited to perform sensitive, wide-field surveys for fast transients compared to their single pixel counterparts. Though the FLAG receiver is more sensitive than all other contemporary PAFs, it lacks their FoV to detect new pulsars and FRBs in the same amount of survey time. The upcoming PAF for the Arecibo telescope which combines the sensitivity of the telescope and the large FoV of other PAFs might be able alleviate these shortcomings. In spite of the lack of discoveries, the pilot survey has proved to be an important step towards identifying the advantages and disadvantages of pulsar surveys with PAF receivers in the future. We plan to undertake a larger survey of the sky with the FLAG that will have much better prospects of finding MSPs and FRBs.

\section*{Acknowledgements}

The authors would like to thank the anonymous referee whose comments vastly improved the quality of the paper. The authors would like to thank Luke Hawkins from the Green Bank Observatory for help during the observations. The authors would like to thank Richard Prestage for his significant contribution to the development of the FLAG receiver. KMR, DRL, DA, NMP, DJP, and MAM acknowledge partial support from National Science Foundation grant AST-1309815. This material is based upon the work supported by National Science Foundation Grant No. 1309832. KMR acknowledges support from European Research Council Horizon 2020 grant (no. 694745) during which part of this work was done. KMR would like to thank Vincent Morello for providing the FFA search code for the survey and Ryan Lynch and Joy Skipper for providing data on the RFI scans close to the observation date.


\bibliographystyle{mnras}
\bibliography{main}

\begin{thebibliography}{}
\makeatletter
\relax
\def\mn@urlcharsother{\let\do\@makeother \do\$\do\&\do\#\do\^\do\_\do\%\do\~}
\def\mn@doi{\begingroup\mn@urlcharsother \@ifnextchar [ {\mn@doi@}
  {\mn@doi@[]}}
\def\mn@doi@[#1]#2{\def\@tempa{#1}\ifx\@tempa\@empty \href
  {http://dx.doi.org/#2} {doi:#2}\else \href {http://dx.doi.org/#2} {#1}\fi
  \endgroup}
\def\mn@eprint#1#2{\mn@eprint@#1:#2::\@nil}
\def\mn@eprint@arXiv#1{\href {http://arxiv.org/abs/#1} {{\tt arXiv:#1}}}
\def\mn@eprint@dblp#1{\href {http://dblp.uni-trier.de/rec/bibtex/#1.xml}
  {dblp:#1}}
\def\mn@eprint@#1:#2:#3:#4\@nil{\def\@tempa {#1}\def\@tempb {#2}\def\@tempc
  {#3}\ifx \@tempc \@empty \let \@tempc \@tempb \let \@tempb \@tempa \fi \ifx
  \@tempb \@empty \def\@tempb {arXiv}\fi \@ifundefined
  {mn@eprint@\@tempb}{\@tempb:\@tempc}{\expandafter \expandafter \csname
  mn@eprint@\@tempb\endcsname \expandafter{\@tempc}}}

\bibitem[\protect\citeauthoryear{{Bannister} et~al.,}{{Bannister}
  et~al.}{2019}]{ban2019}
{Bannister} K.~W.,  et~al., 2019, arXiv e-prints, 1906.11476, \href
  {https://ui.adsabs.harvard.edu/abs/2019arXiv190611476B} {}

\bibitem[\protect\citeauthoryear{{Bates}, {Lorimer}, {Rane}  \&
  {Swiggum}}{{Bates} et~al.}{2014}]{ba14}
{Bates} S.~D.,  {Lorimer} D.~R.,  {Rane} A.,   {Swiggum} J.,  2014, MNRAS, 439,
  2893

\bibitem[\protect\citeauthoryear{{Bhat}, {Cordes}, {Camilo}, {Nice}  \&
  {Lorimer}}{{Bhat} et~al.}{2004}]{bh04}
{Bhat} N.~D.~R.,  {Cordes} J.~M.,  {Camilo} F.,  {Nice} D.~J.,   {Lorimer}
  D.~R.,  2004, ApJ, 605, 759

\bibitem[\protect\citeauthoryear{{Burke-Spolaor} et~al.,}{{Burke-Spolaor}
  et~al.}{2012}]{burke2012}
{Burke-Spolaor} S.,  et~al., 2012, \mn@doi [\mnras]
  {10.1111/j.1365-2966.2012.20998.x}, \href
  {https://ui.adsabs.harvard.edu/abs/2012MNRAS.423.1351B} {423, 1351}

\bibitem[\protect\citeauthoryear{{CHIME/FRB Collaboration} et~al.,}{{CHIME/FRB
  Collaboration} et~al.}{2018}]{CHIME2018}
{CHIME/FRB Collaboration} et~al., 2018, \mn@doi [\apj]
  {10.3847/1538-4357/aad188}, \href
  {http://adsabs.harvard.edu/abs/2018ApJ...863...48C} {863, 48}

\bibitem[\protect\citeauthoryear{{CHIME/FRB Collaboration} et~al.,}{{CHIME/FRB
  Collaboration} et~al.}{2019a}]{chime2019a}
{CHIME/FRB Collaboration} et~al., 2019a, \mn@doi [\nat]
  {10.1038/s41586-018-0867-7}, \href
  {https://ui.adsabs.harvard.edu/abs/2019Natur.566..230C} {566, 230}

\bibitem[\protect\citeauthoryear{{CHIME/FRB Collaboration} et~al.,}{{CHIME/FRB
  Collaboration} et~al.}{2019b}]{chime2019b}
{CHIME/FRB Collaboration} et~al., 2019b, \mn@doi [\nat]
  {10.1038/s41586-018-0864-x}, \href
  {https://ui.adsabs.harvard.edu/abs/2019Natur.566..235C} {566, 235}

\bibitem[\protect\citeauthoryear{{Caleb} et~al.,}{{Caleb} et~al.}{2016}]{ca16}
{Caleb} M.,  et~al., 2016, MNRAS, 458, 718

\bibitem[\protect\citeauthoryear{{Caleb} et~al.,}{{Caleb} et~al.}{2017}]{ca17}
{Caleb} M.,  et~al., 2017, \mn@doi [\mnras] {10.1093/mnras/stx638}, \href
  {https://ui.adsabs.harvard.edu/abs/2017MNRAS.468.3746C} {468, 3746}

\bibitem[\protect\citeauthoryear{{Cameron}, {Barr}, {Champion}, {Kramer}  \&
  {Zhu}}{{Cameron} et~al.}{2017}]{cam17}
{Cameron} A.~D.,  {Barr} E.~D.,  {Champion} D.~J.,  {Kramer} M.,   {Zhu} W.~W.,
   2017, \mn@doi [\mnras] {10.1093/mnras/stx589}, \href
  {http://adsabs.harvard.edu/abs/2017MNRAS.468.1994C} {468, 1994}

\bibitem[\protect\citeauthoryear{{Champion} et~al.,}{{Champion}
  et~al.}{2016}]{ch16}
{Champion} D.~J.,  et~al., 2016, MNRAS

\bibitem[\protect\citeauthoryear{{Chatterjee} et~al.,}{{Chatterjee}
  et~al.}{2017}]{cha17}
{Chatterjee} S.,  et~al., 2017, \nat, 541, 58

\bibitem[\protect\citeauthoryear{{Chawla} et~al.,}{{Chawla}
  et~al.}{2017}]{ch17}
{Chawla} P.,  et~al., 2017, ArXiv e-prints 1701.07457

\bibitem[\protect\citeauthoryear{{Cordes} \& {Lazio}}{{Cordes} \&
  {Lazio}}{2002}]{co02}
{Cordes} J.~M.,  {Lazio} T.~J.~W.,  2002, ArXiv Astrophysics e-prints

\bibitem[\protect\citeauthoryear{{Cordes}, {Wharton}, {Spitler}, {Chatterjee}
  \& {Wasserman}}{{Cordes} et~al.}{2016}]{cor16}
{Cordes} J.~M.,  {Wharton} R.~S.,  {Spitler} L.~G.,  {Chatterjee} S.,
  {Wasserman} I.,  2016, ArXiv e-prints 1605.05890

\bibitem[\protect\citeauthoryear{{Cortes-Medellin}, {Parshley}, {Campbell},
  {Warnick}, {Jeffs}  \& {Ganesh}}{{Cortes-Medellin} et~al.}{2016}]{cortes2016}
{Cortes-Medellin} G.,  {Parshley} S.,  {Campbell} D.~B.,  {Warnick} K.~F.,
  {Jeffs} B.~D.,   {Ganesh} R.,  2016, in Ground-based and Airborne
  Instrumentation for Astronomy VI. p. 99085F, \mn@doi{10.1117/12.2233899}

\bibitem[\protect\citeauthoryear{{Crawford}, {Kaspi}, {Manchester}, {Lyne},
  {Camilo}  \& {D'Amico}}{{Crawford} et~al.}{2001}]{cr2001}
{Crawford} F.,  {Kaspi} V.~M.,  {Manchester} R.~N.,  {Lyne} A.~G.,  {Camilo}
  F.,   {D'Amico} N.,  2001, \mn@doi [\apj] {10.1086/320635}, \href
  {http://adsabs.harvard.edu/abs/2001ApJ...553..367C} {553, 367}

\bibitem[\protect\citeauthoryear{{Deng} et~al.,}{{Deng}
  et~al.}{2017}]{deng2017}
{Deng} X.,  et~al., 2017, \mn@doi [\pasa] {10.1017/pasa.2017.20}, \href
  {https://ui.adsabs.harvard.edu/abs/2017PASA...34...26D} {34, e026}

\bibitem[\protect\citeauthoryear{{Elmer}, {Jeffs}, {Warnick}, {Fisher}  \&
  {Norrod}}{{Elmer} et~al.}{2012}]{elmer2012}
{Elmer} M.,  {Jeffs} B.~D.,  {Warnick} K.~F.,  {Fisher} J.~R.,   {Norrod}
  R.~D.,  2012, \mn@doi [IEEE Transactions on Antennas and Propagation]
  {10.1109/TAP.2011.2173143}, \href
  {https://ui.adsabs.harvard.edu/abs/2012ITAP...60..903E} {60, 903}

\bibitem[\protect\citeauthoryear{{Faucher-Gigu{\`e}re} \&
  {Kaspi}}{{Faucher-Gigu{\`e}re} \& {Kaspi}}{2006}]{fk06}
{Faucher-Gigu{\`e}re} C.-A.,  {Kaspi} V.~M.,  2006, \mn@doi [\apj]
  {10.1086/501516}, \href
  {https://ui.adsabs.harvard.edu/abs/2006ApJ...643..332F} {643, 332}

\bibitem[\protect\citeauthoryear{{Gehrels}}{{Gehrels}}{1986}]{ghe1986}
{Gehrels} N.,  1986, \mn@doi [\apj] {10.1086/164079}, \href
  {https://ui.adsabs.harvard.edu/abs/1986ApJ...303..336G} {303, 336}

\bibitem[\protect\citeauthoryear{{Hewish}, {Bell}, {Pilkington}, {Scott}  \&
  {Collins}}{{Hewish} et~al.}{1968}]{hpb+68}
{Hewish} A.,  {Bell} S.~J.,  {Pilkington} J.~D.~H.,  {Scott} P.~F.,   {Collins}
  R.~A.,  1968, \nat, 217, 709

\bibitem[\protect\citeauthoryear{{James} et~al.,}{{James}
  et~al.}{2019a}]{jam18}
{James} C.~W.,  et~al., 2019a, \mn@doi [\pasa] {10.1017/pasa.2019.1}, \href
  {https://ui.adsabs.harvard.edu/abs/2019PASA...36....9J} {36, e009}

\bibitem[\protect\citeauthoryear{{James}, {Ekers}, {Macquart}, {Bannister}  \&
  {Shannon}}{{James} et~al.}{2019b}]{ja19}
{James} C.~W.,  {Ekers} R.~D.,  {Macquart} J.-P.,  {Bannister} K.~W.,
  {Shannon} R.~M.,  2019b, \mn@doi [\mnras] {10.1093/mnras/sty3031}, \href
  {http://adsabs.harvard.edu/abs/2019MNRAS.483.1342J} {483, 1342}

\bibitem[\protect\citeauthoryear{{Johnston} \& {Karastergiou}}{{Johnston} \&
  {Karastergiou}}{2017}]{kar17}
{Johnston} S.,  {Karastergiou} A.,  2017, \mn@doi [\mnras]
  {10.1093/mnras/stx377}, \href
  {http://adsabs.harvard.edu/abs/2017MNRAS.467.3493J} {467, 3493}

\bibitem[\protect\citeauthoryear{{Keane} et~al.,}{{Keane} et~al.}{2016}]{ke16}
{Keane} E.~F.,  et~al., 2016, \mn@doi [\nat] {10.1038/nature17140}, \href
  {https://ui.adsabs.harvard.edu/abs/2016Natur.530..453K} {530, 453}

\bibitem[\protect\citeauthoryear{{Kramer}}{{Kramer}}{2016}]{kr2016}
{Kramer} M.,  2016, \mn@doi [International Journal of Modern Physics D]
  {10.1142/S0218271816300299}, \href
  {http://adsabs.harvard.edu/abs/2016IJMPD..2530029K} {25, 1630029}

\bibitem[\protect\citeauthoryear{{Levin} et~al.,}{{Levin} et~al.}{2013}]{le13}
{Levin} L.,  et~al., 2013, \mn@doi [\mnras] {10.1093/mnras/stt1103}, \href
  {https://ui.adsabs.harvard.edu/abs/2013MNRAS.434.1387L} {434, 1387}

\bibitem[\protect\citeauthoryear{{Lorimer} \& {Kramer}}{{Lorimer} \&
  {Kramer}}{2004}]{lo04}
{Lorimer} D.~R.,  {Kramer} M.,  2004, {Handbook of Pulsar Astronomy}.
Cambridge University Press

\bibitem[\protect\citeauthoryear{{Lorimer}, {Bailes}, {McLaughlin}, {Narkevic}
  \& {Crawford}}{{Lorimer} et~al.}{2007}]{lo07}
{Lorimer} D.~R.,  {Bailes} M.,  {McLaughlin} M.~A.,  {Narkevic} D.~J.,
  {Crawford} F.,  2007, Science, 318, 777

\bibitem[\protect\citeauthoryear{{Macquart} \& {Ekers}}{{Macquart} \&
  {Ekers}}{2018}]{mac2018}
{Macquart} J.-P.,  {Ekers} R.~D.,  2018, \mn@doi [\mnras]
  {10.1093/mnras/stx2825}, \href
  {http://adsabs.harvard.edu/abs/2018MNRAS.474.1900M} {474, 1900}

\bibitem[\protect\citeauthoryear{{Malenta} et~al.,}{{Malenta}
  et~al.}{2018}]{mal2018}
{Malenta} M.,  et~al., 2018, in {Weltevrede} P.,  {Perera} B.~B.~P.,  {Preston}
  L.~L.,   {Sanidas} S.,  eds,  IAU Symposium Vol. 337, Pulsar Astrophysics the
  Next Fifty Years. pp 370--371, \mn@doi{10.1017/S1743921317008511}

\bibitem[\protect\citeauthoryear{{Manchester} et~al.,}{{Manchester}
  et~al.}{2001}]{ma01a}
{Manchester} R.~N.,  et~al., 2001, MNRAS, 328, 17

\bibitem[\protect\citeauthoryear{{McLaughlin}}{{McLaughlin}}{2013}]{mcl13}
{McLaughlin} M.~A.,  2013, \mn@doi [Classical and Quantum Gravity]
  {10.1088/0264-9381/30/22/224008}, \href
  {http://adsabs.harvard.edu/abs/2013CQGra..30v4008M} {30, 224008}

\bibitem[\protect\citeauthoryear{{McLaughlin} et~al.,}{{McLaughlin}
  et~al.}{2006}]{mc06}
{McLaughlin} M.~A.,  et~al., 2006, \nat, 439, 817

\bibitem[\protect\citeauthoryear{{Morgan}, {Fisher}  \& {Castro}}{{Morgan}
  et~al.}{2013}]{morgan2013}
{Morgan} M.~A.,  {Fisher} J.~R.,   {Castro} J.~J.,  2013, \mn@doi [\pasp]
  {10.1086/671349}, \href
  {https://ui.adsabs.harvard.edu/abs/2013PASP..125..695M} {125, 695}

\bibitem[\protect\citeauthoryear{{Obrocka}, {Stappers}  \&
  {Wilkinson}}{{Obrocka} et~al.}{2015}]{ob2015}
{Obrocka} M.,  {Stappers} B.,   {Wilkinson} P.,  2015, \mn@doi [\aap]
  {10.1051/0004-6361/201425538}, \href
  {http://adsabs.harvard.edu/abs/2015A%26A...579A..69O} {579, A69}

\bibitem[\protect\citeauthoryear{{Parent} et~al.,}{{Parent}
  et~al.}{2018}]{pa2018}
{Parent} E.,  et~al., 2018, \mn@doi [\apj] {10.3847/1538-4357/aac5f0}, \href
  {http://adsabs.harvard.edu/abs/2018ApJ...861...44P} {861, 44}

\bibitem[\protect\citeauthoryear{{Petroff} et~al.,}{{Petroff}
  et~al.}{2015a}]{pe15}
{Petroff} E.,  et~al., 2015a, MNRAS, 447, 246

\bibitem[\protect\citeauthoryear{{Petroff} et~al.,}{{Petroff}
  et~al.}{2015b}]{pe15b}
{Petroff} E.,  et~al., 2015b, \mnras, 451, 3933

\bibitem[\protect\citeauthoryear{{Petroff} et~al.,}{{Petroff}
  et~al.}{2016}]{pet16}
{Petroff} E.,  et~al., 2016, \mn@doi [\pasa] {10.1017/pasa.2016.35}, \href
  {http://adsabs.harvard.edu/abs/2016PASA...33...45P} {33, e045}

\bibitem[\protect\citeauthoryear{{Phinney} \& {Kulkarni}}{{Phinney} \&
  {Kulkarni}}{1994}]{ku1994}
{Phinney} E.~S.,  {Kulkarni} S.~R.,  1994, \mn@doi [\araa]
  {10.1146/annurev.aa.32.090194.003111}, \href
  {http://adsabs.harvard.edu/abs/1994ARA%26A..32..591P} {32, 591}

\bibitem[\protect\citeauthoryear{{Rajwade} et~al.,}{{Rajwade}
  et~al.}{2018a}]{raj2018}
{Rajwade} K.~M.,  et~al., 2018a, in {Weltevrede} P.,  {Perera} B.~B.~P.,
  {Preston} L.~L.,   {Sanidas} S.,  eds,  IAU Symposium Vol. 337, Pulsar
  Astrophysics the Next Fifty Years. pp 398--399 (\mn@eprint {arXiv}
  {1710.09650}), \mn@doi{10.1017/S1743921317009012}

\bibitem[\protect\citeauthoryear{{Rajwade}, {Chennamangalam}, {Lorimer}  \&
  {Karastergiou}}{{Rajwade} et~al.}{2018b}]{raj2018b}
{Rajwade} K.,  {Chennamangalam} J.,  {Lorimer} D.,   {Karastergiou} A.,  2018b,
  \mn@doi [\mnras] {10.1093/mnras/sty1695}, \href
  {http://adsabs.harvard.edu/abs/2018MNRAS.479.3094R} {479, 3094}

\bibitem[\protect\citeauthoryear{{Ransom}}{{Ransom}}{2008}]{ran2008}
{Ransom} S.~M.,  2008, in {Bassa} C.,  {Wang} Z.,  {Cumming} A.,   {Kaspi}
  V.~M.,  eds,  American Institute of Physics Conference Series Vol. 983, 40
  Years of Pulsars: Millisecond Pulsars, Magnetars and More. pp 415--423
  (\mn@eprint {arXiv} {0710.3626}), \mn@doi{10.1063/1.2900267}

\bibitem[\protect\citeauthoryear{{Ransom}}{{Ransom}}{2011}]{ran2011}
{Ransom} S.,  2011, {PRESTO: PulsaR Exploration and Search TOolkit},
  Astrophysics Source Code Library (\mn@eprint {ascl} {1107.017})

\bibitem[\protect\citeauthoryear{{Roshi} et~al.,}{{Roshi}
  et~al.}{2018}]{rosh2018}
{Roshi} D.~A.,  et~al., 2018, \mn@doi [\aj] {10.3847/1538-3881/aab965}, \href
  {http://adsabs.harvard.edu/abs/2018AJ....155..202R} {155, 202}

\bibitem[\protect\citeauthoryear{Roshi, Shillue  \& Fisher}{Roshi
  et~al.}{2019}]{roshi2019}
Roshi D.~A.,  Shillue W.,   Fisher J.~R.,  2019, \mn@doi [{IEEE} Transactions
  on Antennas and Propagation] {10.1109/tap.2019.2899046}, 67, 3011

\bibitem[\protect\citeauthoryear{Shannon et~al.,}{Shannon
  et~al.}{2018}]{sh2018}
Shannon R.~M.,  et~al., 2018, \mn@doi [Nature] {10.1038/s41586-018-0588-y},
  562, 386

\bibitem[\protect\citeauthoryear{{Staelin}}{{Staelin}}{1969}]{st69}
{Staelin} D.~H.,  1969, IEEE Proceedings, \href
  {http://adsabs.harvard.edu/abs/1969IEEEP..57..724S} {57, 724}

\bibitem[\protect\citeauthoryear{{Surnis} et~al.,}{{Surnis}
  et~al.}{2019}]{greenburst}
{Surnis} M.~P.,  et~al., 2019, arXiv e-prints, \href
  {https://ui.adsabs.harvard.edu/abs/2019arXiv190305573S} {p. arXiv:1903.05573}

\bibitem[\protect\citeauthoryear{{Tendulkar} et~al.,}{{Tendulkar}
  et~al.}{2017}]{te17}
{Tendulkar} S.~P.,  et~al., 2017, \apjl, 834, L7

\bibitem[\protect\citeauthoryear{{Thornton} et~al.,}{{Thornton}
  et~al.}{2013}]{th13}
{Thornton} D.,  et~al., 2013, Science, 341, 53

\bibitem[\protect\citeauthoryear{{Warnick}, {Carter}, {Webb}, {Landon}, {Elmer}
   \& {Jeffs}}{{Warnick} et~al.}{2011}]{warnick2011}
{Warnick} K.~F.,  {Carter} D.,  {Webb} T.,  {Landon} J.,  {Elmer} M.,   {Jeffs}
  B.~D.,  2011, \mn@doi [IEEE Transactions on Antennas and Propagation]
  {10.1109/TAP.2011.2122223}, \href
  {https://ui.adsabs.harvard.edu/abs/2011ITAP...59.1876W} {59, 1876}

\makeatother
\end{thebibliography}




\bsp	
\label{lastpage}
\end{document}